\documentclass[11pt]{article}
\def\vec#1{\ifmmode
\mathchoice{\mbox{\boldmath$\displaystyle\bf#1$}}
{\mbox{\boldmath$\textstyle\bf#1$}}
{\mbox{\boldmath$\scriptstyle\bf#1$}}
{\mbox{\boldmath$\scriptscriptstyle\bf#1$}}\else
{\mbox{\boldmath$\bf#1$}}\fi}

\usepackage{placeins}

\begin{document}


\begin{flushleft}
\end{flushleft}

\vspace{2 cm}

\begin{center}
{\LARGE Testing the approximations described in \\}
{\LARGE ``Asymptotic formulae for likelihood-based tests of new physics''}
\end{center}

\vspace{3 cm}

\begin{center}
Eric Burns, Wade Fisher
\end{center}

\vspace{0.5 cm}

\noindent 
Department of Physics and Astronomy, Michigan State University, East
Lansing, MI 48825 

\vspace{3 cm}

\begin{abstract}
``Asymptotic formulae for likelihood-based tests of new physics" presents a mathematical formalism for a new approximation for hypothesis testing in high energy physics. The approximations are designed to greatly reduce the computational burden for such problems. We seek to test the conditions under which the approximations described remain valid. To do so, we perform parallel calculations for a range of scenarios and compare the full calculation to the approximations to determine the limits and robustness of the approximation. We compare this approximation against values calculated with the {\tt Collie} framework, which for our analysis we assume produces true values.
\end{abstract}

\vspace{1 cm}
\noindent
Keywords:  systematic uncertainties,
profile likelihood,
hypothesis test,
confidence interval, 
frequentist methods, 
asymptotic methods,
asimov data set,
{\tt Collie},
AWW approximation

\clearpage

\tableofcontents

\clearpage

\section{Introduction}
\label{sec:intro}

One of the primary goals in experimental particle physics is the search for new particles. In order to determine whether or not a particle has been discovered statistical hypothesis tests are used. The probability of finding an outcome as extreme as the one observed can be compared to a predetermined threshold to ascertain whether or not discovery has occured. 

\noindent Unfortunately, due to the sheer magnitude of the amount of data involved in the search for the new particles, determining probabilities is often computationally intensive. In this paper we examine the approximation presented in ``Asymptotic formulae for likelihood-based tests of new physics," to find the limits of its applicability. This approximation is evaluated to determine when it successfully reproduces the results from a full semi-frequentist computation with no approximations (Section \ref{sec:Pdata}). Conclusions based on these findings are presented in Section \ref{sec:Conc}.

Presented below is the necessary prerequisite knowledge; this includes general statistics (Section \ref{sec:Hy}), such as hypothesis testing and the likelihood ratio (Section \ref{sec:LiFu}), as well as an explanation of how these techniques are used in particle physics (Section \ref{sec:PaPhSt}). We then explain the mathematical basis for the Asimov data set based upon results from Wilks and Wald (Section \ref{sec:AsAp}), as given by the authors of \cite{Asym}. The Asimov data set is a representative set of values that theoretically represents the true parameters of the full ensemble. This set contains represents an ensemble of simulated data; later it is described in greater depth (Section \ref{sec:AsAp}). Henceforth the three approximations together will be abbreivated as the AWW approximation, an acronym of their names. This allows us to examine the possibility that the approximation generates valid parameters. The approximation, the full mathematical formalism and subsequent evidence are presented in (arXiv:1007.1727v2), upon which our explanation and formalism are based \cite{Asym}.

\section{Mathematical Formalism}
\label{sec:formalism}

Presented here are some basic statistical principles, such as hypothesis testing and test statistics, as well as more complex ideas like the likelihood function and it's application to binned data. This section ends with a brief overview of statistical methods used in particle physics.

\subsection{Basic Statistics}
\label{sec:Hy}

A hypothesis is a suggested solution to explain a given phenomenom. One often compares the validity of two hypotheses through statistical testing, where one decides whether a given null hypothesis, $H_0$, should be rejected in favor of the alternate hypotheses, $H_1$. In particle physics the null hypothesis typically contains all known processes and the alternate hypothesis may also contain a new process or particle. Meaning, the null hypothesis would be background-only and the alternate hypothesis would then be signal-plus-background.

A test statistic is a function of the sample and assumed to be a numerical summary of the data that can be used to reject, or fail to reject, a hypothesis. This can be done by calculating the probability of obtaining a test statistic as extreme as the one observed, which is called a $p$-value. This represents the level of agreement between the data and a single hypothesis. The $p$-value can be measured against a significance level $\alpha$, defined as the critical $p$-value; i.e. $p$ must be less than or equal to $\alpha$ to reject a given hypothesis. 

\noindent The $p$-value can also be converted to a standardized value, such as a $Z$-score, the number of standard deviations a datum is from the mean; $Z$ is given as a function of $p$ by

\begin{equation}
\label{eq:DefineZScore}
Z(p) = \Phi^{-1}(1-p) \,,
\end{equation}

\noindent where $\Phi^{-1}(p)=\sqrt{2}Erf^{-1}(2p-1)$, the quantile of the standard Gaussian\footnote{Erf is the error function. $Erf^{-1}(z)=\displaystyle\sum_{k=0}^{\infty} \frac {c_k}{2k+1} (\frac {\sqrt{\pi}}{2}z)^{2k+1}$ \\ \indent \indent \indent where $c_k=\displaystyle\sum_{m=0}^{k-1} \frac {c_mc_{k-1-m}} {(m+1)(2m+1)}=\{1,1,\frac {7}{6},\frac {127}{90}\},...$}. At $\alpha=0.05$, a commonly used signficance level, the $Z$-score is equal to 1.64 for a one-sided test; a one-sided test is used when the critical outcomes capable of rejecting a hypothesis occur on only one side of the distribution. Because we can distinguish between positive and negative fluctuations in our tests we use a one-sided test, with a 95\% confidence level (CL) exclusion.

\subsection{The Likelihood Function and Maximization}
\label{sec:LiFu}

The likelihood of a given observation given a set of parameters is equal to the probability of a set of parameter values given an observation. 

\noindent Consider a set of N observables, contained in $\vec{x}=(x_1,...,x_N)$, described by probability distribution function (p.d.f.) $f(\vec{x};\vec{\theta})$, where $\vec{\theta}=(\theta_1,...,\theta_n)$ are the unknown parameters, which also known as the nuisance parameters. Assuming statistical independence between the measurements $x_i$, then the likelihood function L(\vec{\theta}) is

\begin{equation}
\label{eq:LikelihoodFunction}
L(\vec{\theta}) = \prod_{i=1}^N{f(x_i;\vec{\theta})}.
\end{equation}

The $\vec{\theta}$ values that maximize this function are denoted $\hat{\vec{\theta}}$. In order to find the maximum likelihood (ML) estimators one can solve the formula \cite{Nakamura}

\begin{equation}
\label{eq:MLEstimators}
\frac{\partial \ln L}{\partial \theta_i}=0,\indent i = 1,...,n.
\end{equation}

\noindent The covariance matrix of the ML estimators, $V_{ij}=\mbox{cov}[\hat{\theta_i},\hat{\theta_j}]$ can be used to estimate the standard deviation, $\sigma$. We can find this by first finding the inverse covariance matrix, which can be approximated as

\begin{equation}
\label{eq:InverseCovarianceMatrix}
(\hat{V^{-1}})_{ij}=-\frac{\partial^2 \ln L}{\partial \theta_i 
\partial \theta_j}\bigg|_{\vec{\hat{\theta}}},
\end{equation}

\noindent and then invert the resulting matrix to find the standard deviation. This is also known as the curvature matrix, and can only be used when the positive and negative deviations are equal. 

\subsection{Likelihood Approximation for Binned Data}
\label{sec:LiApBiDa}

If a sample size is large it is often easier bin the data into a histogram. This results in a vector $\vec{n}=(n_1,...,n_N)$ with expectation value 
$\vec{\nu}=E[\vec{n}]$ and p.d.f. $f(\vec{n},\vec{\nu})$. Maximizing the likelihood ratio is equivalent to minimizing the quantity $-2\ln\lambda(\vec{\theta})$. For independent, Poisson distributed $n_i$ this quantity is \cite{Baker}

\begin{equation}
\label{eq:BiLiAp}
-2\ln\lambda(\vec{\theta})=2\sum_{i=1}^N\left[\nu_i(\vec{\theta})
-n_i+n_i \ln \frac {n_i} {\nu_i(\vec{\theta})}\right],
\end{equation}

\noindent where the last term is zero when $n_i=0$.  According to Wilks' theorem, for sufficiently large samples that meet certain regularity conditions, the minimum of Eq.~(\ref{eq:BiLiAp}) follows a $\chi^2$ distribution, allowing the usage of goodness-of-fit tests \cite{Wilks}. 

\subsection{Particle Physics Statistics}
\label{sec:PaPhSt}

This subsection describes how the forementioned statistical principles are often applied in particle physics. In particle physics a $Z$-score greater than or equal to 5, or $p = 2.87 \times 10^{-7}$ for a one-sided tail, is usually required for discovery, which results from the rejection of the background-only hypothesis.

For binned data with a histogram of variable $x$ and information $\vec{n}=(n_1,....,n_N)$, the expectation value

\begin{equation}
\label{eq:ExpectationValue}
E[n_i] = \mu s_i+b_i,
\end{equation}

\noindent where $\mu$ is the signal strength, and $s_i$ and $b_i$ are the mean number of entries in the $i$th bin, meaning \cite{Asym}

\begin{eqnarray}
\label{eq:Si}
s_i = s_{\rm tot} \int_{{\rm bin} \, i} f_{s}(x; \vec{\theta}_{s}) \, dx \,, 
\\*[0.3 cm]
\label{eq:Bi}
b_i = b_{\rm tot} \int_{{\rm bin} \, i} f_{b}(x; \vec{\theta}_{b}) \, dx \,.
\end{eqnarray}

\noindent Here $f_s(x;\vec{\theta}_s$) and $f_b(x;\vec{\theta}_b$) are the p.d.f.s of the variable $x$ for signal and background events respectively. The signal strength is equal to zero for the background-only hypothesis and one for the nominal signal hypothesis. Henceforth, $\vec{\theta}$ contains all nuisance parameters, i.e. $\vec{\theta}=(\vec{\theta_s},\vec{\theta_b},b_{tot})$; $s_{tot}$ is not contained in $\vec{\theta}$ because it's value is fixed by the prediction from the nominal signal hypothesis.

\noindent One can create a control sample that measures only background events, with information contained in histogram $\vec{m}=(m_1,...,m_M)$ the expectation value of $m_i$ is

\begin{equation}
\label{eq:Mi}
E[m_i]=u_i(\vec{\theta}),
\end{equation}

\noindent where $u_i$ is dependent on the nuisance parameters. The purpose of the control sample is to add useful constraints to the nuisance parameters. 

Using the signal-plus-background and background-only information, the likelihood function can be written as a product of two Poisson probabilities

\begin{equation}
\label{eq:DoPoLiFu}
L(\mu,\vec{\theta})=\prod_{j=1}^N \frac{ (\mu s_{j} + b_{j} )^{n_{j}} }
{ n_{j}! } e^{- (\mu s_{j} + b_{j}) }   \;\; \prod_{k=1}^M 
\frac{ u_k^{m_{k}}} { m_{k}! } \, e^{- u_k }  \;.
\end{equation}

\noindent The test statistic we are interested in is $\gamma=-2\ln\lambda(\mu)$, where 

\begin{equation}
\label{eq:PrLiRa}
\lambda(\mu)=\frac {L(\mu,\hat{\hat{\vec{\theta}}})}
{L(\hat{\mu},\hat{\vec{\theta}})}
\end{equation}

\noindent is the profile likelihood ratio. Here $\hat{\hat{\vec{\theta}}}$ denotes the conditional maximum-likelihood estimator for the specified $\mu$; $\hat{\mu}$ and $\hat{\vec{\theta}}$ are the unconditional maximum-likelihood estimators.

Assigning our value as $\gamma$', we can calculate the $p$-value from

\begin{equation}
\label{eq:pVal}
p(\mu)=\int_\gamma^{\infty} f(\gamma|\mu)dt,
\end{equation}

\noindent where $f(\gamma|\mu)$ is the p.d.f. of $\gamma$ for the given signal strength $\mu$ \cite{Asym}.

\section{The Asimov Data Set Approximation}
\label{sec:AsAp}

The conditional definition of the Asimov data set is that when one uses it to evaluate the estimators for all parameters one obtains the true parameter values, i.e. it represents the maximum likelihood for the parent p.d.f. In order to test if the Asimov condition holds one can use the generic likelihood function Eq.~(\ref{eq:LikelihoodFunction}). Using the simplified notation $\nu_i=\mu's_i+b_i$, and setting $\theta_0=\mu$, then Eq.~(\ref{eq:MLEstimators}) becomes

\begin{equation}
\label{eq:MLAsimov}
\frac {\partial \ln L} {\partial \theta_j} = 
\displaystyle\sum_{i=1}^N\left(\frac {n_i}{\nu_i}-1\right)
\frac {\partial \nu_i}{\partial \theta_j} +
\displaystyle\sum_{i=1}^M\left(\frac {m_i} {u_i} -1\right)
\frac {\partial u_i} {\partial \theta_j} = 0.
\end{equation}

\noindent If $n_{i,A}=E(n_i)$ and $m_{i,A}=E[m_i]$, where the subscript A denotes Asimov values, then the Asimov condition is met. We cannot calculate the Asimov likelihood $L_A$ because it contains factorial dependence on Asimov values that can be non-integer. However, these factorials are canceled in the Asimov profile likelihood ratio

\begin{equation}
\label{eq:AsimovLambda} 
\lambda_{\rm A}(\mu) 
= \frac{ L_{\rm A}( \mu, \hat{\hat{\vec{\theta}}} ) } { L_{\rm A}(\hat{\mu},
\hat{\vec{\theta}}) } = \frac{ L_{\rm A}( \mu,
\hat{\hat{\vec{\theta}}} ) } { L_{A}(\mu^{\prime}, \vec{\theta} ) } \;,
\end{equation}

\noindent where the substitution in the denominator of the final equality is allowed by the definition of the Asimov data set \cite{Asym}.

\subsection{The Wald Equation}
\label{sec:WaldEquation}

Suppose we have a test with strength parameter $\mu$ and the data is distributed by strength parameter $\mu^{'}$, then according to Wald \cite{Wald}

\begin{equation}
\label{eq:wald}
-2 \ln \lambda(\mu)
= \frac{(\mu - \hat{\mu})^2}{\sigma^2} + {\cal  O}(1/\sqrt{N}) \;,
\end{equation}

\noindent where N is the sample size and $\hat{\mu}$ is a Gaussian distribution with mean $\mu'$. Here $\sigma$ is found using the covariance matrix.

Substituting the Asimov data set with strength parameter $\mu'$ into the Wald approximation equation, it follows from Eq.~(\ref{eq:wald}) that

\begin{equation}
\label{eq:WaldAsimov}
-2\ln \lambda_A(\mu)\approx \frac {(\mu-\mu^{'})^2} {\sigma^2}
\end{equation}

\noindent for large samples. We provide an alternate way to find the standard deviation via the Asimov data set, defining $q_{\mu,A}=-2\ln \lambda_A(\mu)$, 

\begin{equation}
\label{eq:AsimovStdDev}
\sigma_A^2=\frac {(\mu-\mu')^2} {q_{\mu,A}}.
\end{equation}

\noindent To find the median exclusion significance assuming there is no signal $\mu'=0$, Eq.~(\ref{eq:AsimovStdDev}) reduces to

\begin{equation}
\label{eq:AsimovMuPrimeZero}
\sigma_A^2=\frac {\mu^2} {q_{\mu,A}}.
\end{equation}

\noindent Similarly for the case of discovery where $\mu=0$, Eq.~(\ref{eq:AsimovStdDev}) is

\begin{equation}
\label{eq:AsimovMuZero}
\sigma_A^2=\frac {\mu'^{2}} {q_{\mu,A}}.
\end{equation}

\subsection{The Tevatron Test Statistic}
\label{eq:TeTeSt}
The test statistic

\begin{equation}
\label{eq:TevTestStat}
q=-2\ln \frac {L_{s+b}}{L_b},
\end{equation}

\noindent is often used in analyses at the Fermilab Tevatron Collider. Here $L_{s+b}$ is the nominal signal model with strength parameter $\mu=1$, and $L_b$ is the background-only hypothesis with $\mu=0$. Rewriting Eq.~(\ref{eq:TevTestStat}),

\begin{equation}
\label{TevTestStatSimp1}
q=-2\ln \frac {L(\mu=1,\hat{\hat{\vec{\theta}}}(1))}
{L(\mu=0,\hat{\hat{\vec{\theta}}}(0))}
= -2 \ln \lambda(1) + 2\ln \lambda(0).
\end{equation}

\noindent If the Wald appromixation holds, then

\begin{equation}
\label{TevTestStatSimp2}
q= \frac {(\hat{\mu}-1)^2} {\sigma^2} - \frac {\hat{\mu}^2} {\sigma^2}
= \frac {1-2\hat{\mu}} {\sigma^2}.
\end{equation}

\noindent Since $\hat{\mu}$ is Gaussian and $q$ is dependent on $\hat{\mu}$ then $q$ is also Gaussian. Therefore, the expectation value and standard deviation of $q$ are \cite{Asym}

\begin{eqnarray}
\label{eq:E[q]}
E[q] = \frac {1-2\mu} {\sigma^2},
\\*[0.3 cm]
\label{eq:sigma[q]}
\sigma[q] = \frac {2} {\sigma(\mu)}.
\end{eqnarray}

ASince $q$ is Gaussian we can use the cumulative distribution function\footnote{For a normal variable with mean $\mu$, variance $\sigma^2$ and observation x the cumulative distribution function is $\Phi(\frac {x-\mu} {\sigma}) =\frac {1} {2} [1 + erf(\frac{x-\mu} {\sigma/\sqrt{2}})]$} to determine the $p$-value. Plugging in what we know of the signal strengths of the two hypotheses, as well as the mean and standard deviation of q,\footnote{The original paper contains confusing notation and a substitution error in their derivation; the formulas presented here are correct.}

\begin{eqnarray}
\label{eq:psb}
p_{s+b}=\int_{q_{obs}}^{\infty}f(q|s+b)dq=
1-\Phi\left(\frac {q_{obs}+1/\sigma_{s+b}^2}{2/\sigma_{s+b}}\right),
\\*[0.3 cm]
\label{eq:pb}
p_b=\int_{-\infty}^{q_{obs}}f(q|b)dq=
\Phi\left(\frac {q_{obs}-1/\sigma_b^2}{2/\sigma_b^2}\right).
\end{eqnarray}

\section{Pseudo-data Tests}
\label{sec:Pdata}
In order to test if the AWW approximation reproduces the real distributions of $\gamma$ we created a set of test data, applied various systematic uncertainties and compared with the values produced by the {\tt Collie} framework. We calculate the signal strength required to achieve a given significance level in both models and compare.

The pseudo-data generated has least likelihood ratios similar to a set of Tevatron data by construction, and is displayed in Fig.~\ref{fig:BaaScSi}. We define the data as equal to the background before systematic uncertainties.

\setlength{\unitlength}{1.0 cm}
\renewcommand{\baselinestretch}{0.9}
\begin{figure}[htbp]
\begin{picture}(10.0,5)
\put(0,0)
{\includegraphics{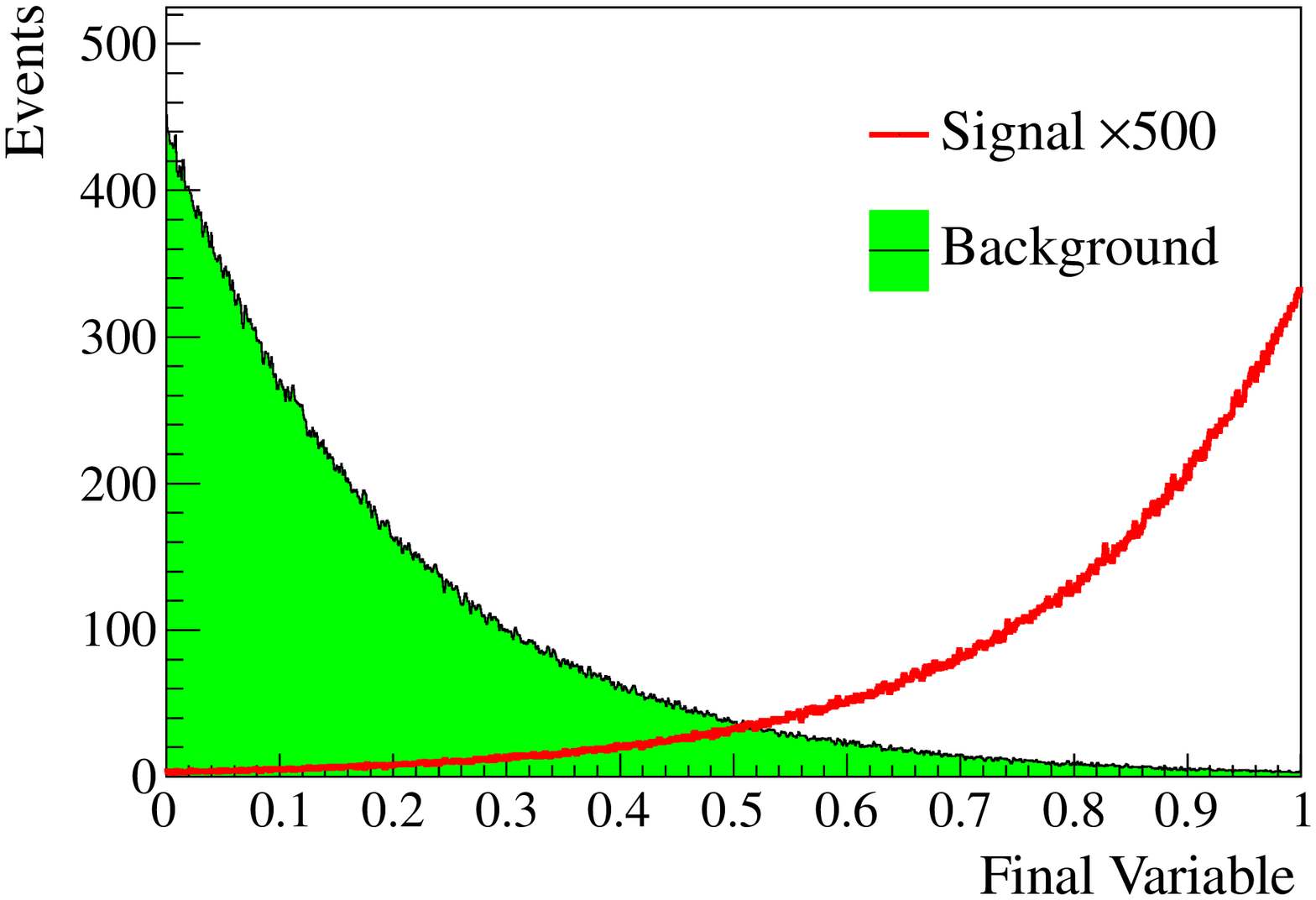}}
\put(8,0)
{\includegraphics{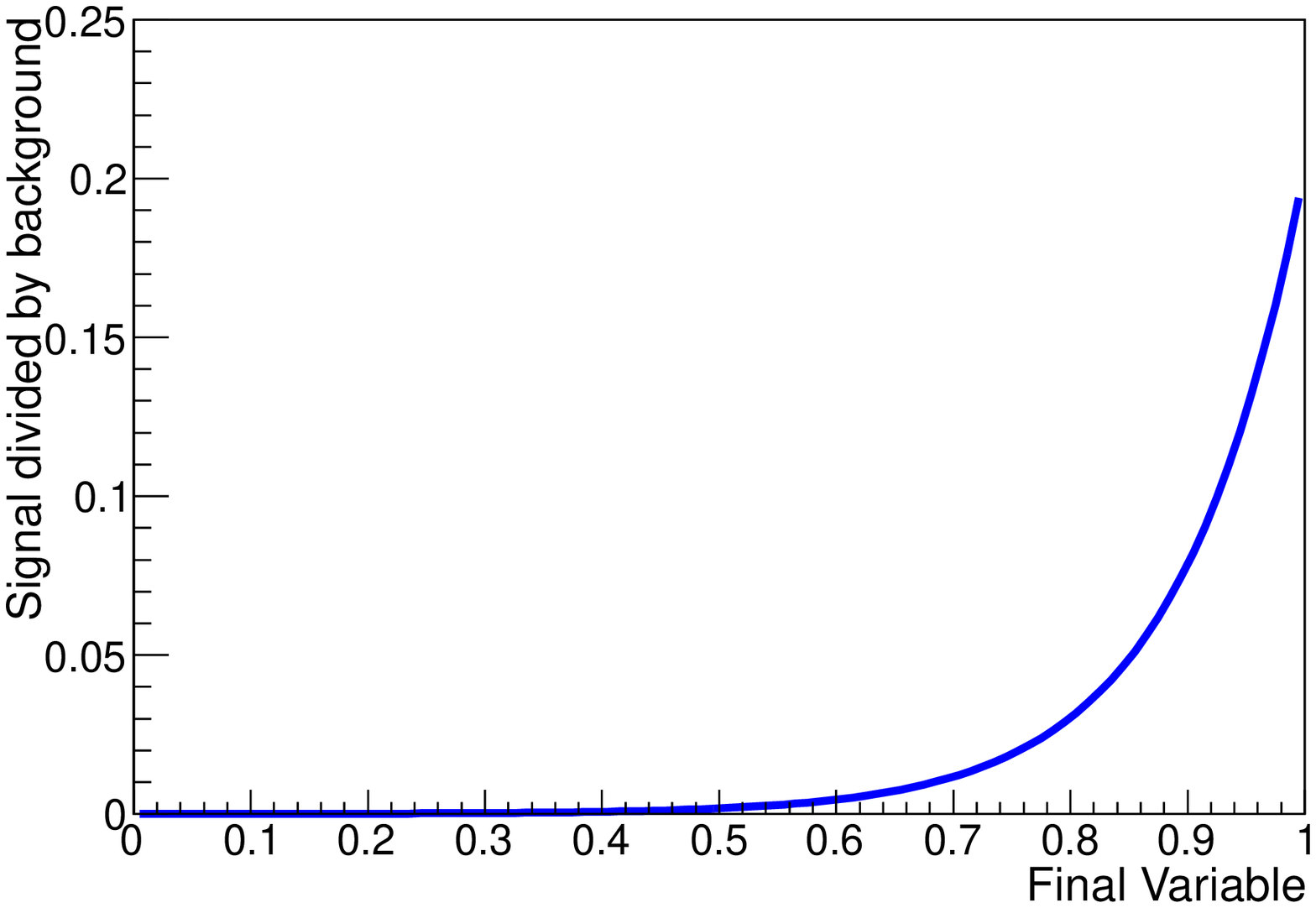}}
\end{picture}
\caption{\small On the left is a plot of the test input data generated in root with 1,000,000 events placed into 1500 bins. The background was filled with an exponential decay function, $e^{\tau x}$, with $\tau=0.203$ and weight 135000/event count, and the signal was filled by a mirrored exponential decay function, $1-e^{\tau x}$, with $\tau=0.215$ and weight 214/event count. For ease of intepretation the signal is displayed with a scale factor of 500 and the y-axis is linear. This histogram has 1500 bins and 1,000,000 events, which are used in this paper unless stated otherwise. On the right we display the ratio of signal over background.}
\label{fig:BaaScSi}
\end{figure}
\renewcommand{\baselinestretch}{1}
\small\normalsize

The {\tt Collie} software suite generates semi-frequentist confidence intervals with an output designed for Root \cite{Fisher}. Here we will consider the {\tt Collie} confidence level value to be true for the sake of evaluating the Asimov conditions. {\tt Collie} also outputs the observed, signal plus background, and background-only least likelihood ratios, which are used to calculate the AWW approximation.

From Eq.~(\ref{eq:AsimovMuPrimeZero}), with $\mu=1$ from the nominal signal hypotheses we have

\begin{equation}
\label{eq:PDVar}
\sigma_{s+b,A}^2=\frac {1} {q_{s+b}}.
\end{equation}
 
\noindent Substituting this value into Eq.~(\ref{eq:psb}) 

\begin{equation}
\label{eq:PDpSB}
P_{s+b}=1-\Phi\left(\frac {q_{obs} + q_{s+b}} {2\sqrt{q_{s+b}}}\right),
\end{equation}

\noindent which provides a simple calculation of the AWW approximation using the {\tt Collie} output. We report results in terms of a ratio; this ratio is always the approximation value divided by the {\tt Collie} value. We keep this standard because the Wald approximation should result in underestimation, thus the ratio should stay below one.

\FloatBarrier

\subsection{Background-only Rate Systematic Uncertainty}
\label{BaonSy}

The first systematic uncertainty applied was a rate systematic uncertainty on only the background. Our results are plotted in Fig.~(\ref{fig:TempBaaScSi}). As expected we see no discrepancy when there is no uncertainty, i.e. when the background rate systematic uncertainty is set at 0\%, meaning the data and background are equal.

\setlength{\unitlength}{1.0 cm}
\renewcommand{\baselinestretch}{0.9}
\begin{figure}[htbp]
\begin{picture}(10.0,10)
\put(0,5)
{\includegraphics{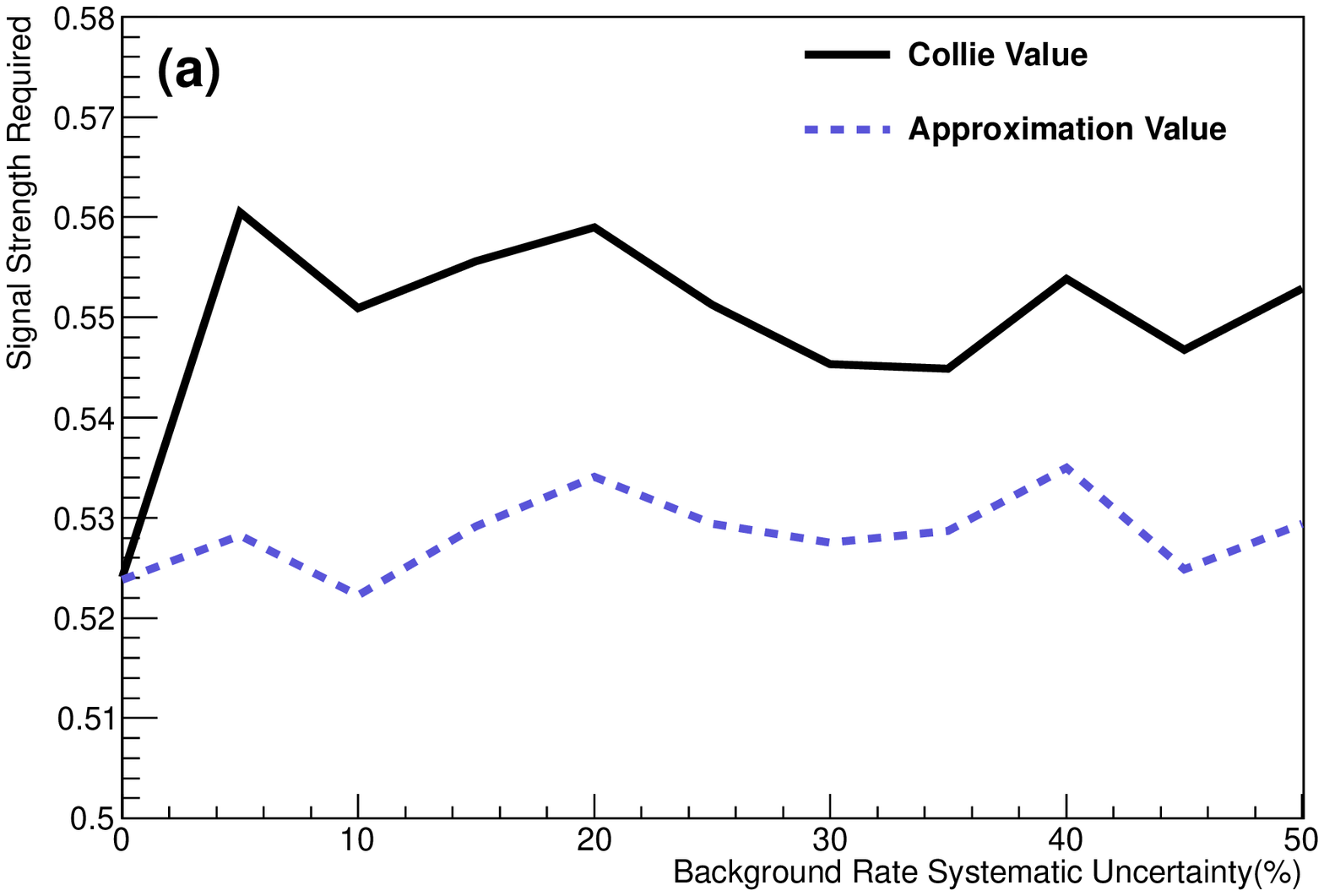}}
\put(8,5)
{\includegraphics{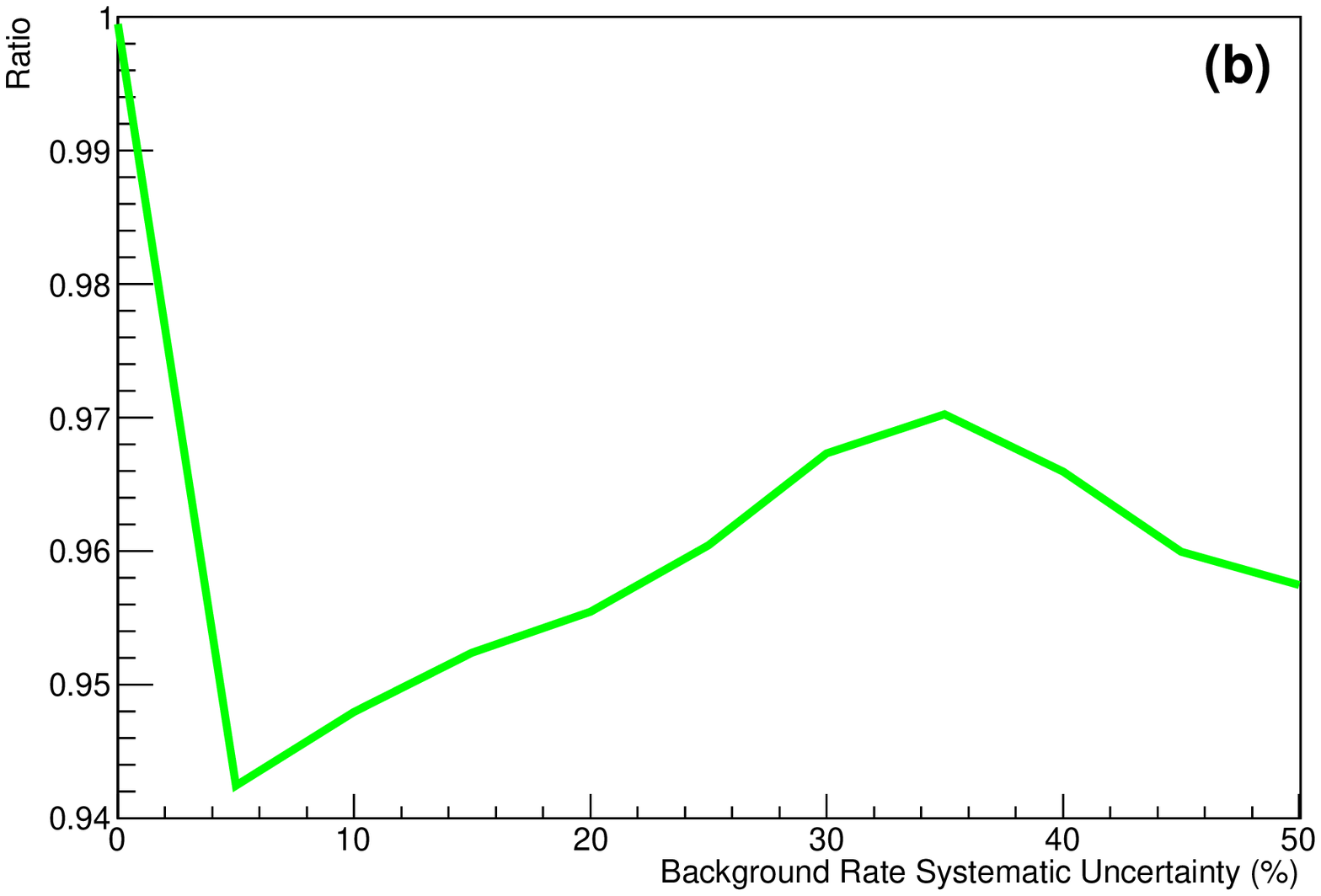}}
\put(0,0)
{\includegraphics{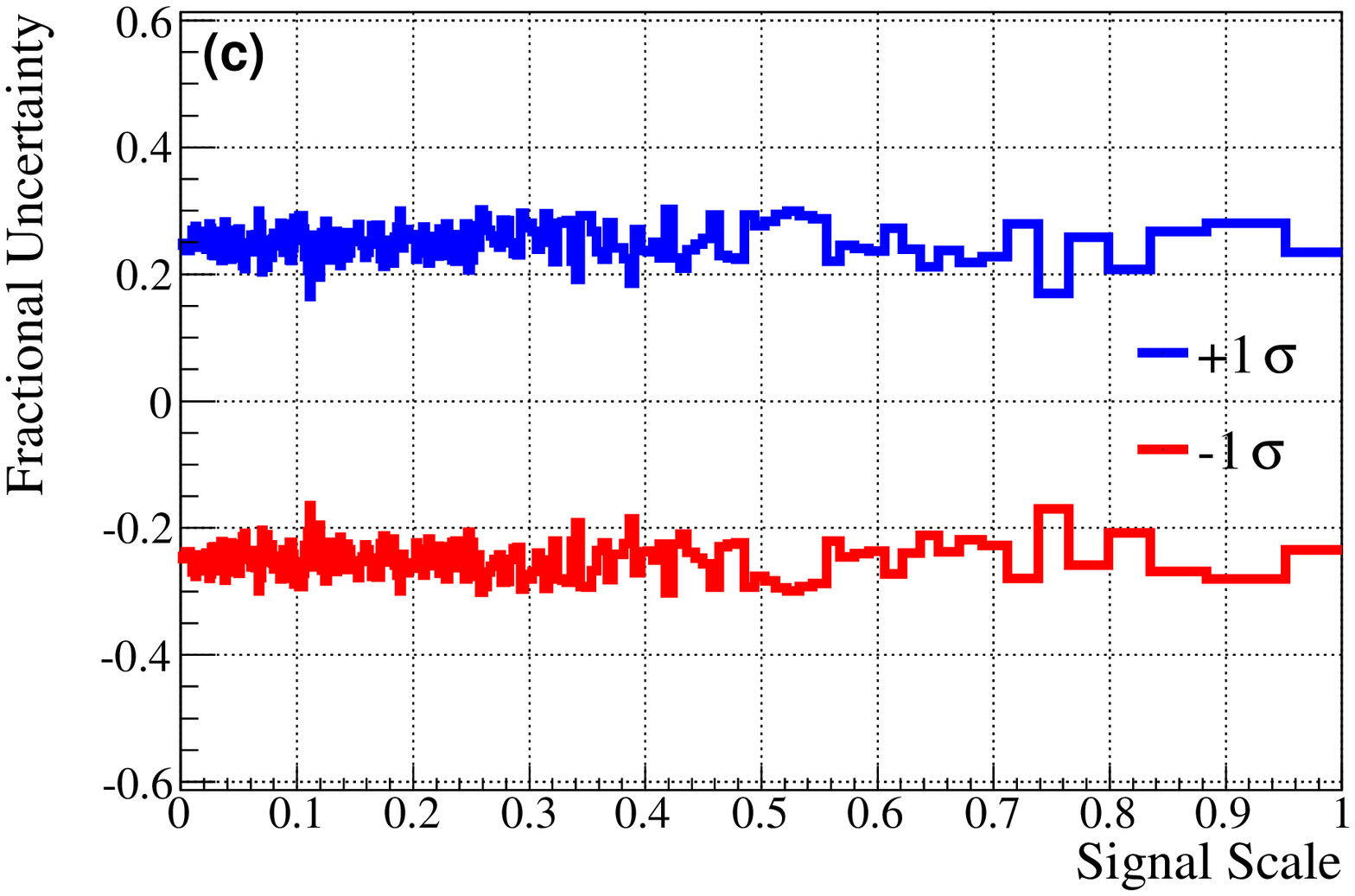}}
\put(8.0,0.4){\makebox(8,4)[b]{\begin{minipage}[b]{7cm}
\protect\caption{{\footnotesize This experiment measured the two methods against each other while they accounted for a background-only rate systematic uncertainty that varied from zero to fifty percent in five percent increments. (a) Shows the signal scale necessary to achieve the CL for both methods, (b) shows the ratio of the approximation value over the {\tt Collie} value. (c) shows the fractional uncertainty of the background at 25\% uncertainty in the rate systematic uncertainty.}
\protect\label{fig:TempBaaScSi}}
\end{minipage}}}

\end{picture}

\label{fig:background}
\end{figure}
\renewcommand{\baselinestretch}{1}
\small\normalsize

\FloatBarrier

\noindent As we apply the rate systematic uncertainty we get up to around 5\% deviation from the ``true" value, as well as no obvious trend as a function of systematic uncertainty percent. Therefore the AWW approximation is valid.


\subsection{Signal and Background Rate Systematic Uncertainties}
\label{SiBaSS}

Perhaps the most striking results were the three dimensional plots where the axes in the horizontal plane represent the percent rate systematic uncertainties of the signal and background histograms. We created plots of both uncorrelated and correlated systematic uncertainties. No systematic uncertainty plots are shown as they are equivalent to the systematic uncertainty plot in the background rate systematic uncertainty, only now applied to signal as well as background.

\setlength{\unitlength}{1.0 cm}
\renewcommand{\baselinestretch}{0.8}
\begin{figure}[htbp]
\begin{picture}(10.0,5.5)
\put(1,0){\includegraphics{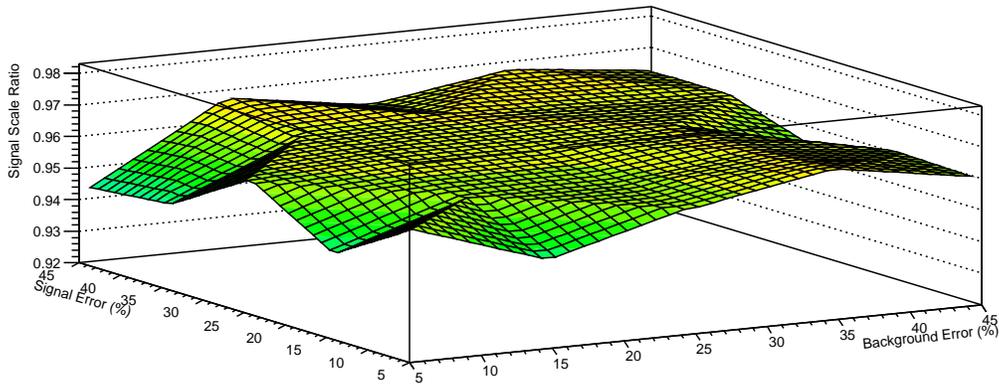}}
\end{picture}
\caption{\footnotesize The horizontal plane contains the axes for the background and signal rate systematic uncertainties percent, varying from 5\% to 45\% in 10\% increments. The variable we are interested in is the ratio of the AWW approximation over the {\tt Collie} value, which is presented in the z-axis. The color is a gradient and the scale is held in this manner for comparison to the correlated plot.}
\label{fig:GUnc}
\end{figure}
\renewcommand{\baselinestretch}{1}
\small\normalsize

\setlength{\unitlength}{1.0 cm}
\renewcommand{\baselinestretch}{0.8}
\begin{figure}[htbp]
\begin{picture}(10.0,5.5)
\put(1,0){\includegraphics{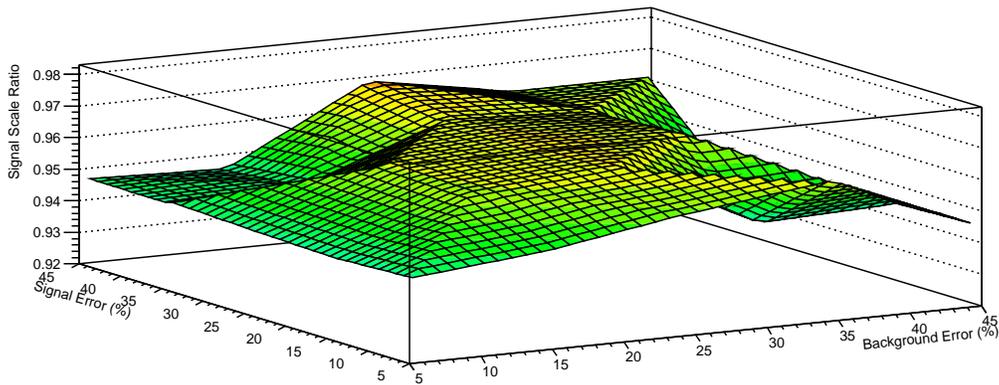}}
\end{picture}
\caption{\footnotesize The axes, scale and confidence level are equivalent to that of Fig.~(\ref{fig:GUnc}). These plots appear fairly similar}
\label{fig:GCor}
\end{figure}
\renewcommand{\baselinestretch}{1}
\small\normalsize

\noindent Fig.~(\ref{fig:GUnc}) displays the uncorrelated data set and Fig.~(\ref{fig:GCor}) the correlated. With relatively flat signal scale ratios at the C.L. we conclude that the AWW approximation is valid for these systematic uncertainties.

\FloatBarrier

\subsection{Asymmetric Gaussian ``Flat'' Systematic Uncertainties}
\label{sec:DiFS}

For the next experiment we ran two tests with a flat systematic uncertainties with a discontinuity at the center. Fig.~(\ref{fig:APlots}) shows the way in which {\tt Collie} approximates a solution for an asymmetric Gaussian as well as the systematic uncertainty itself. The first test had the positive systematic uncertainty constant and the negative varied, while the second reversed the roles.

\setlength{\unitlength}{1.0 cm}
\renewcommand{\baselinestretch}{0.9}
\begin{figure}[htbp]
\begin{picture}(10.0,5.5)
\put(0,.2)
{\includegraphics{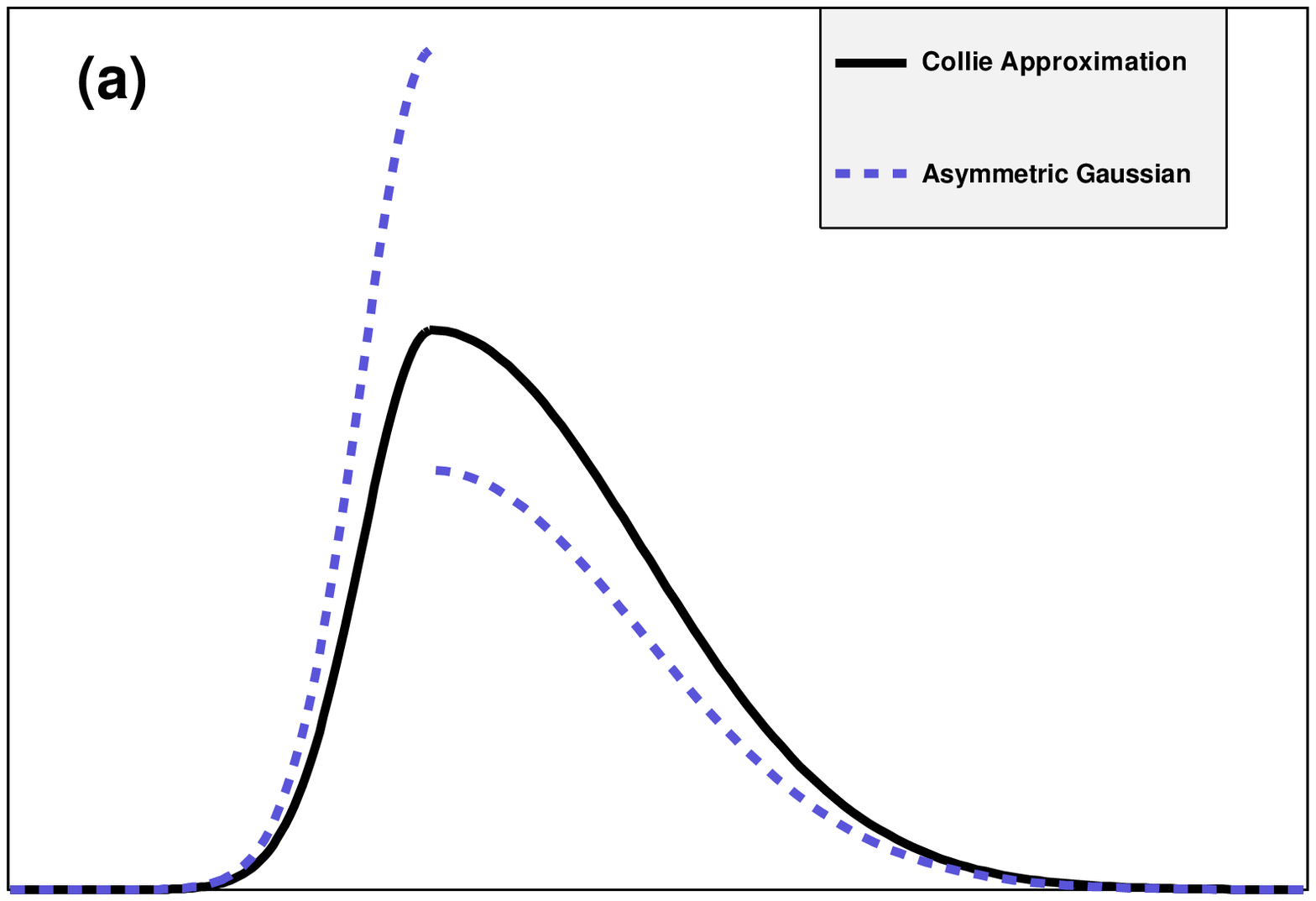}}
\put(8,0)
{\includegraphics{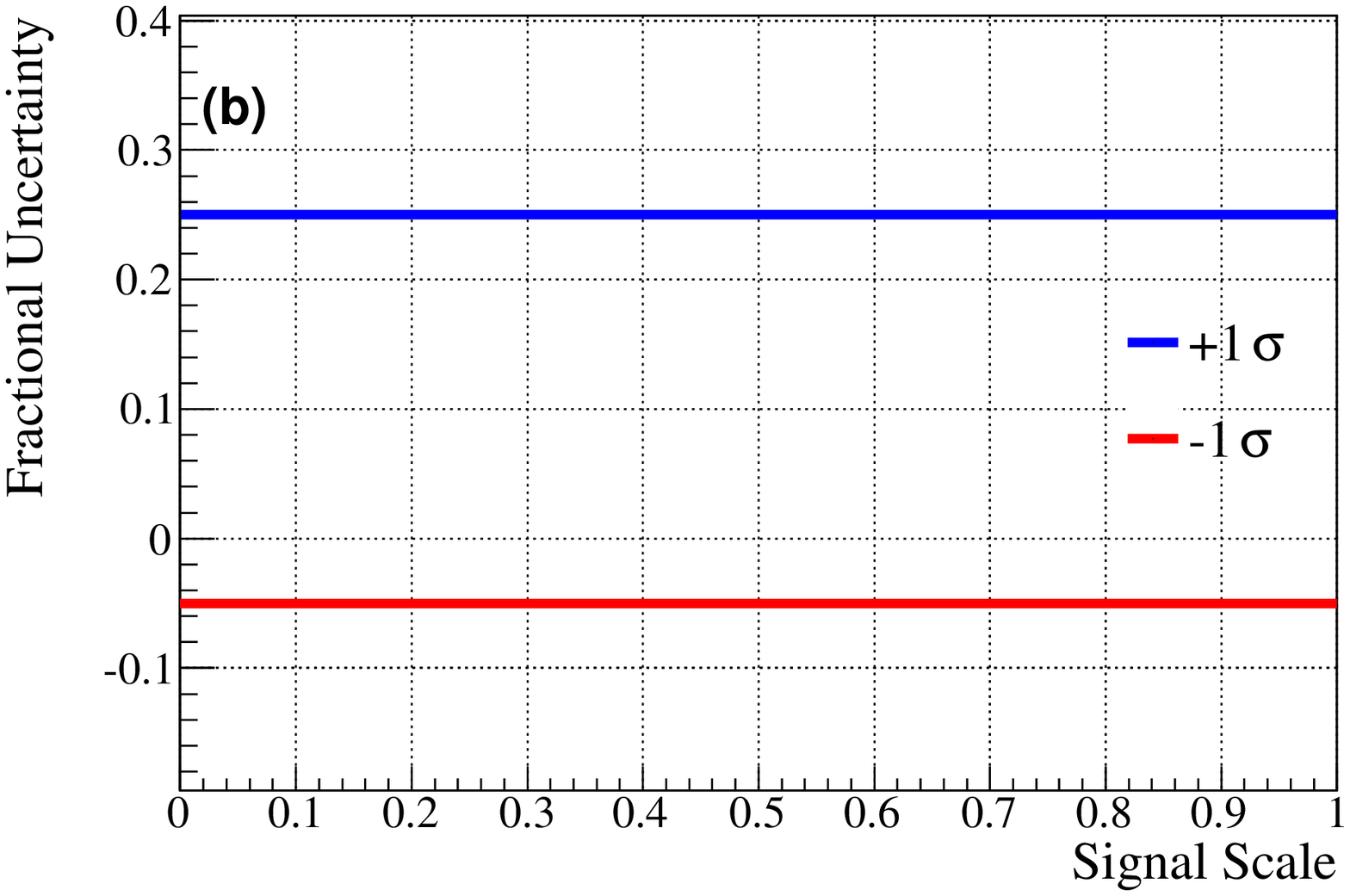}}
\end{picture}
\caption{\small The plot in (a) shows the collie approxmation to an asymmetric Gaussian. (b) shows the flat systematic uncertainty at negative fluctuations of 5\% and positive fluctuations of 10\%.}
\label{fig:APlots}
\end{figure}
\renewcommand{\baselinestretch}{1}
\small\normalsize

\setlength{\unitlength}{1.0 cm}
\renewcommand{\baselinestretch}{0.9}
\begin{figure}[htbp]
\begin{picture}(10.0,5)
\put(0,0)
{\includegraphics{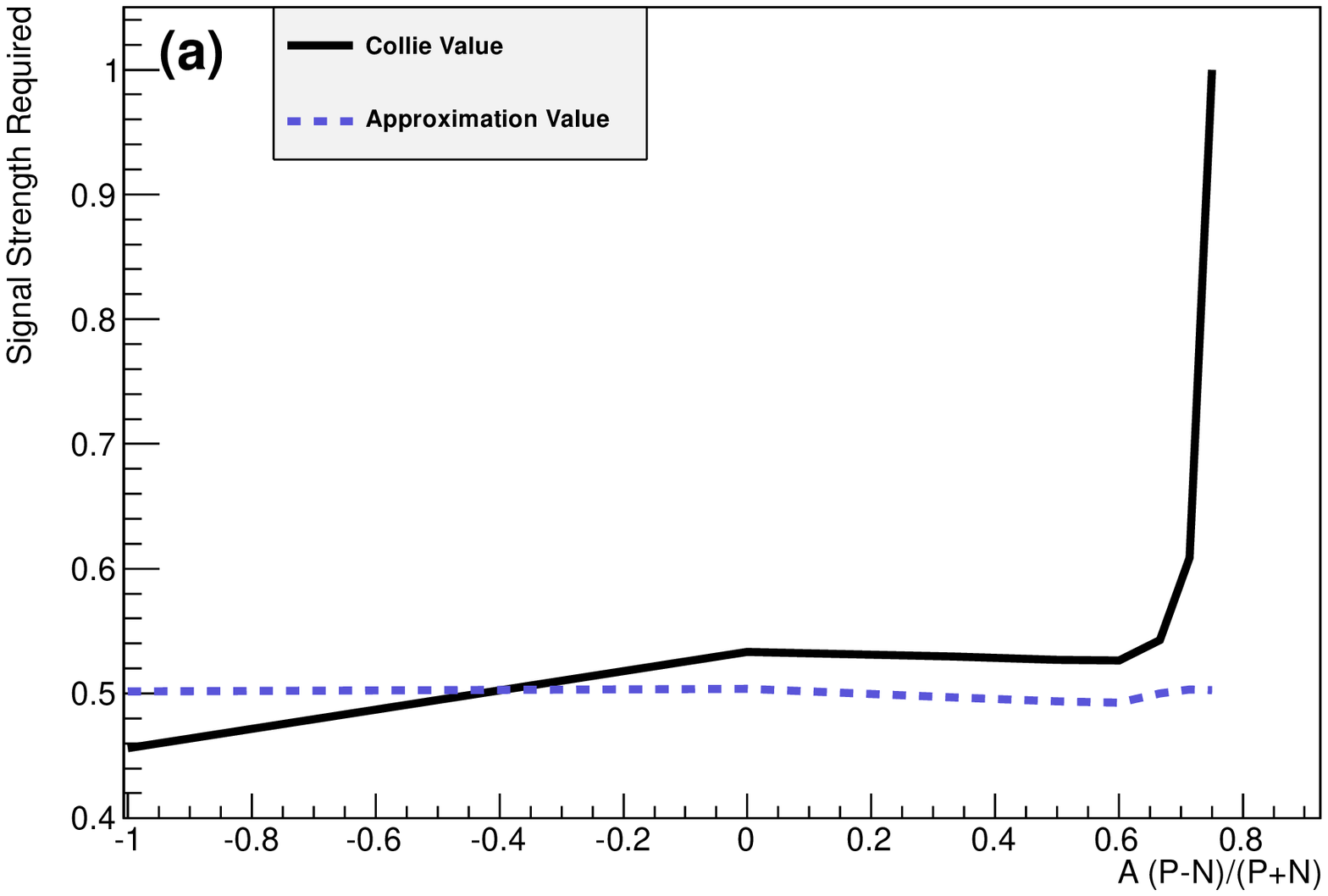}}
\put(8,0)
{\includegraphics{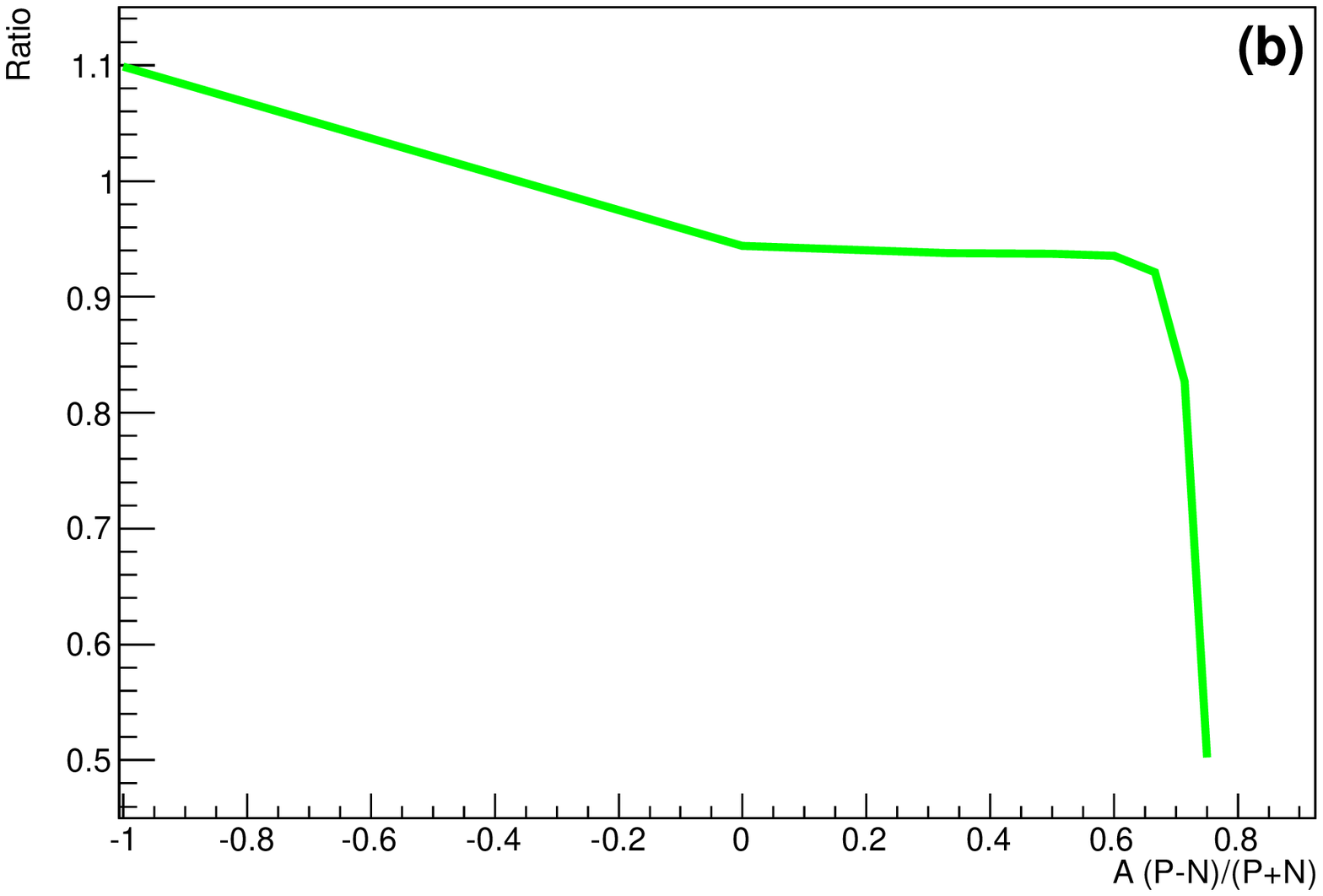}}
\end{picture}
\caption{\small Here the negative side of the flat systematic uncertainty was held at 5\% while the positive varied from 0\% to 35\%. (a) The signal scale necessary required is 90\% confidence, (b) shows the ratio}
\label{fig:ANeg}
\end{figure}
\renewcommand{\baselinestretch}{1}
\small\normalsize

\setlength{\unitlength}{1.0 cm}
\renewcommand{\baselinestretch}{0.9}
\begin{figure}[htbp]
\begin{picture}(10.0,5)
\put(0,0)
{\includegraphics{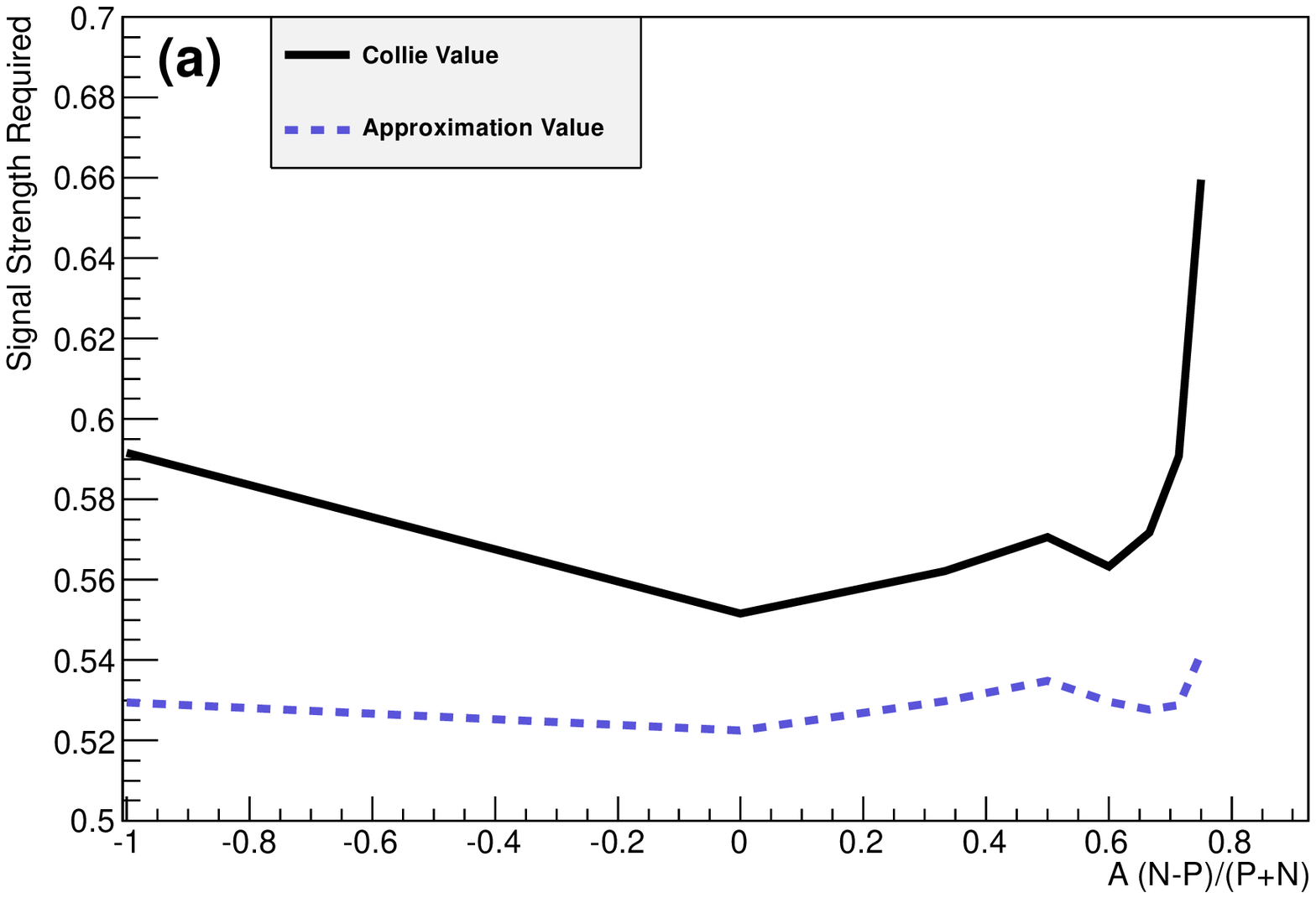}}
\put(8,0)
{\includegraphics{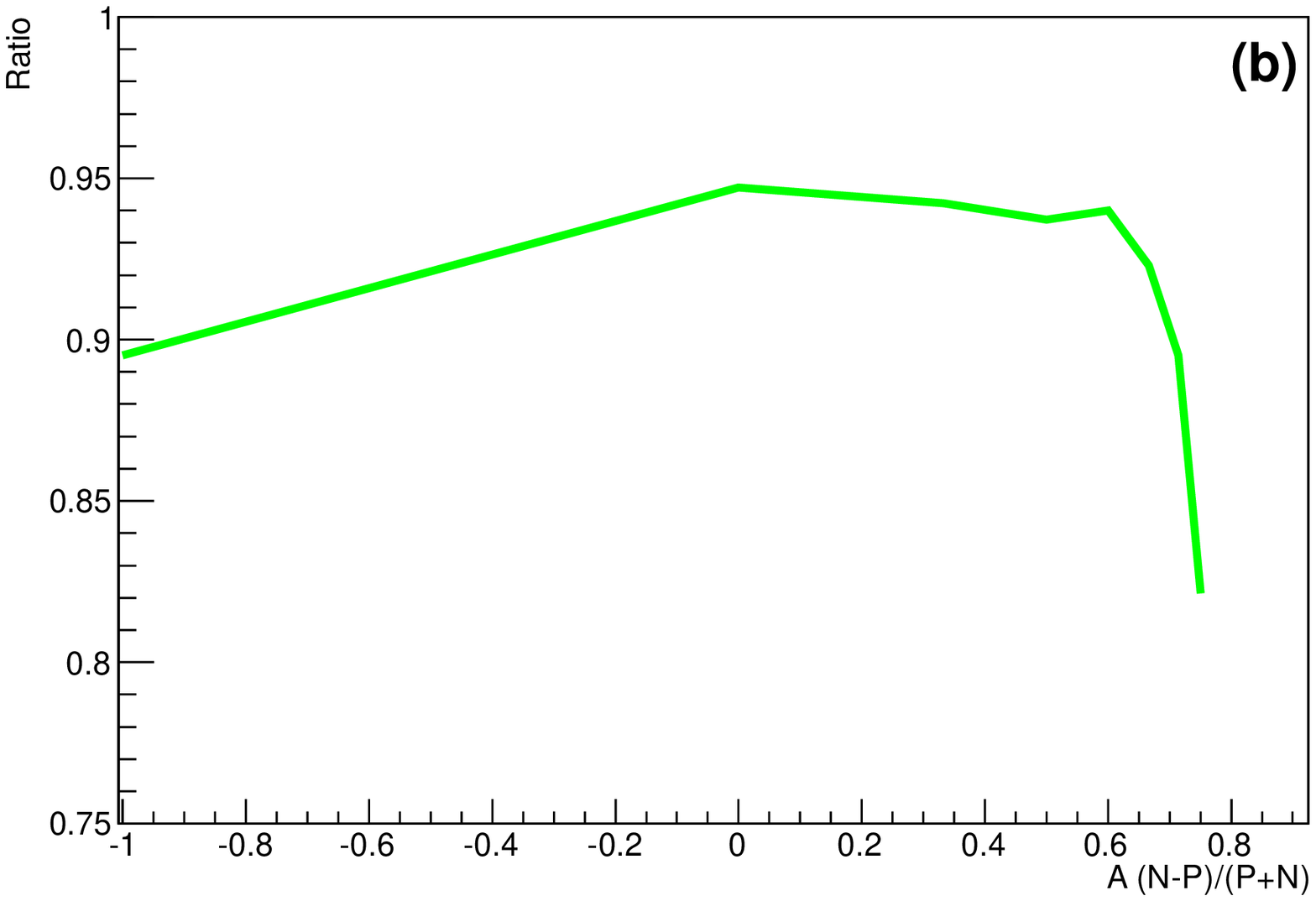}}
\end{picture}
\caption{\small This displays the same information as Fig.~(\ref{fig:ANeg}) except for this experiment the positive component is held fixed while the
negative is varied.}
\label{fig:APos}
\end{figure}
\renewcommand{\baselinestretch}{1}
\small\normalsize

One notable difference here from the other tests is that we had to use the observed {\tt Collie} confidence level instead of the calculated median, which results in slightly greater random variability. This is due to the systematic uncertainty being non-Gaussian.

\noindent Both sets were run from 0\% to 50\% on the uncertainty that varies, but are only plotted up to 35\%. This is because the data at and above 35\% return unusable values due to a failure in the AWW approximation. This occurs because at this level and type of systematic uncertainty the histograms are no longer Gaussian. When there is 5\% negative and no positive uncertainty the AWW approximation overestimates the value. Other than this, at low uncertainty differences the AWW approximation is still valid, however above 35\% on the varying systematic uncertainty it is invalid as the model breaks down.

\subsection{Uncertainty on Background Shape}
\label{BaTaVa}

Next, we tested the resilience of the AWW approximation against deviations in the tau of the exponential decay function of the background. The initial $\tau$ value we used, 0.203, was chosen in order to simulate the least likelihood ratio values found in a set of real Tevatron data (this is also true for the case of the signal tau formula, where $\tau = 0.215$). Fig.~(\ref{fig:tau}) displays these findings. 

\setlength{\unitlength}{1.0 cm}
\renewcommand{\baselinestretch}{0.9}
\begin{figure}[htbp]
\begin{picture}(10.0,10)
\put(0,5)
{\includegraphics{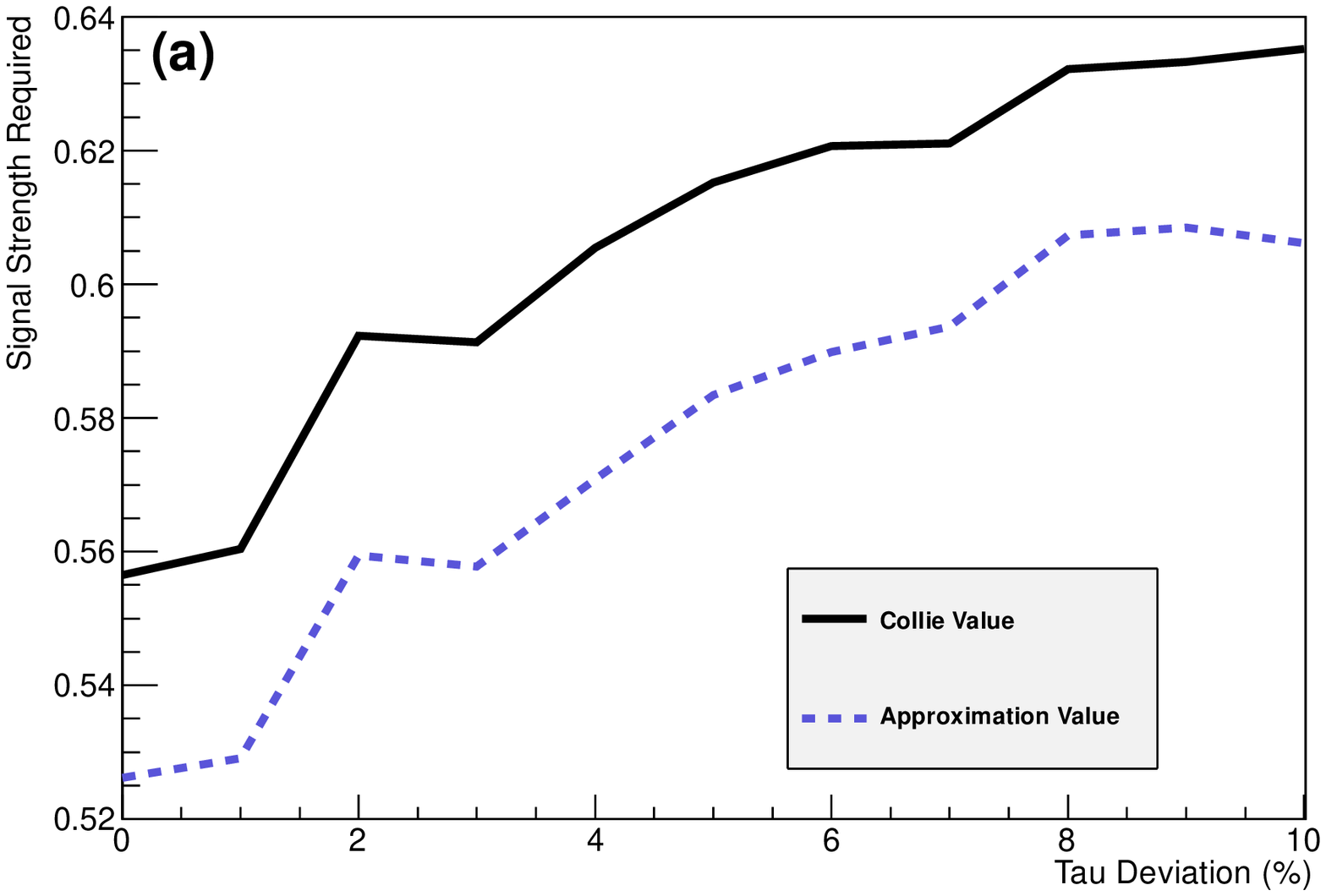}}
\put(8,5)
{\includegraphics{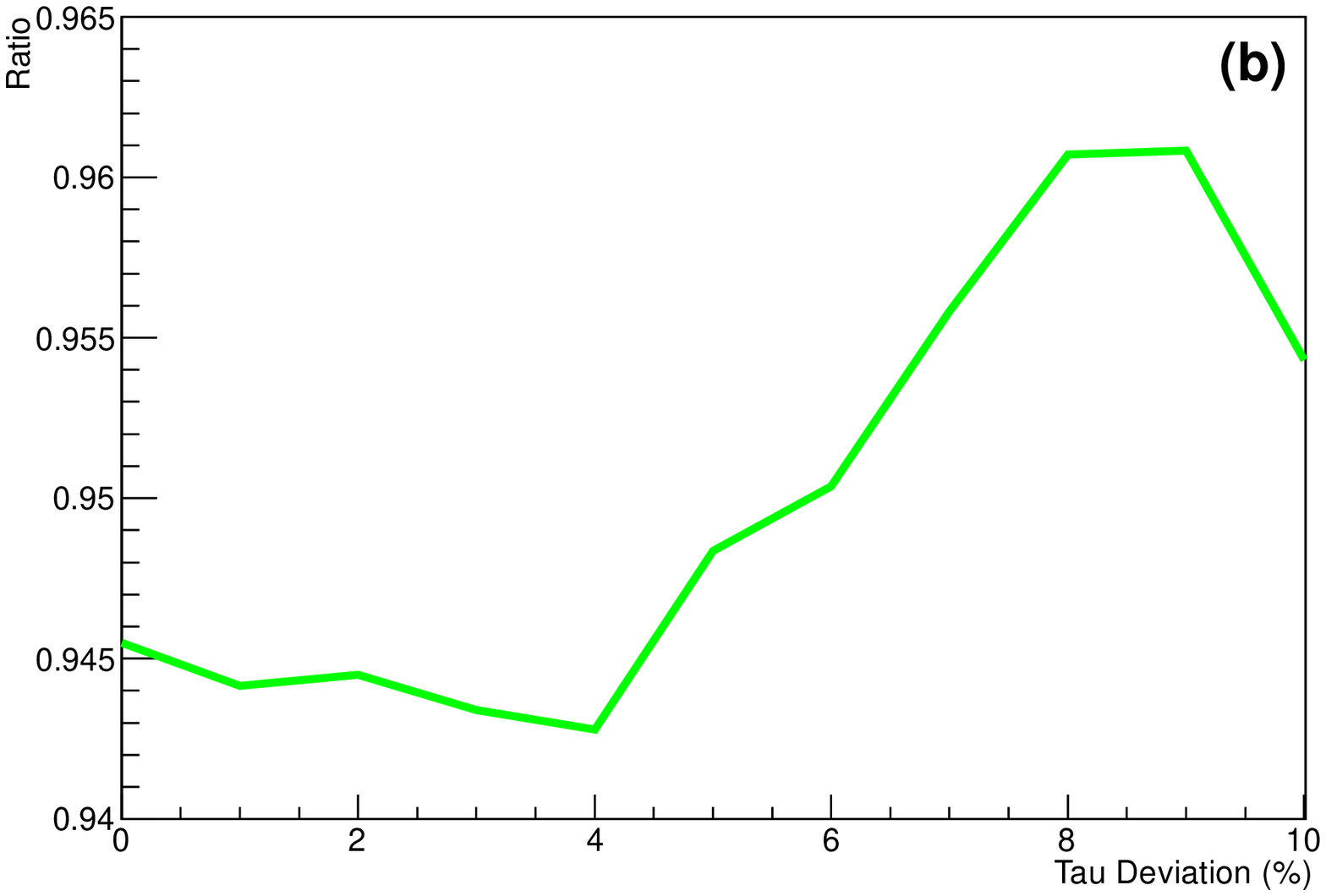}}
\put(0,0)
{\includegraphics{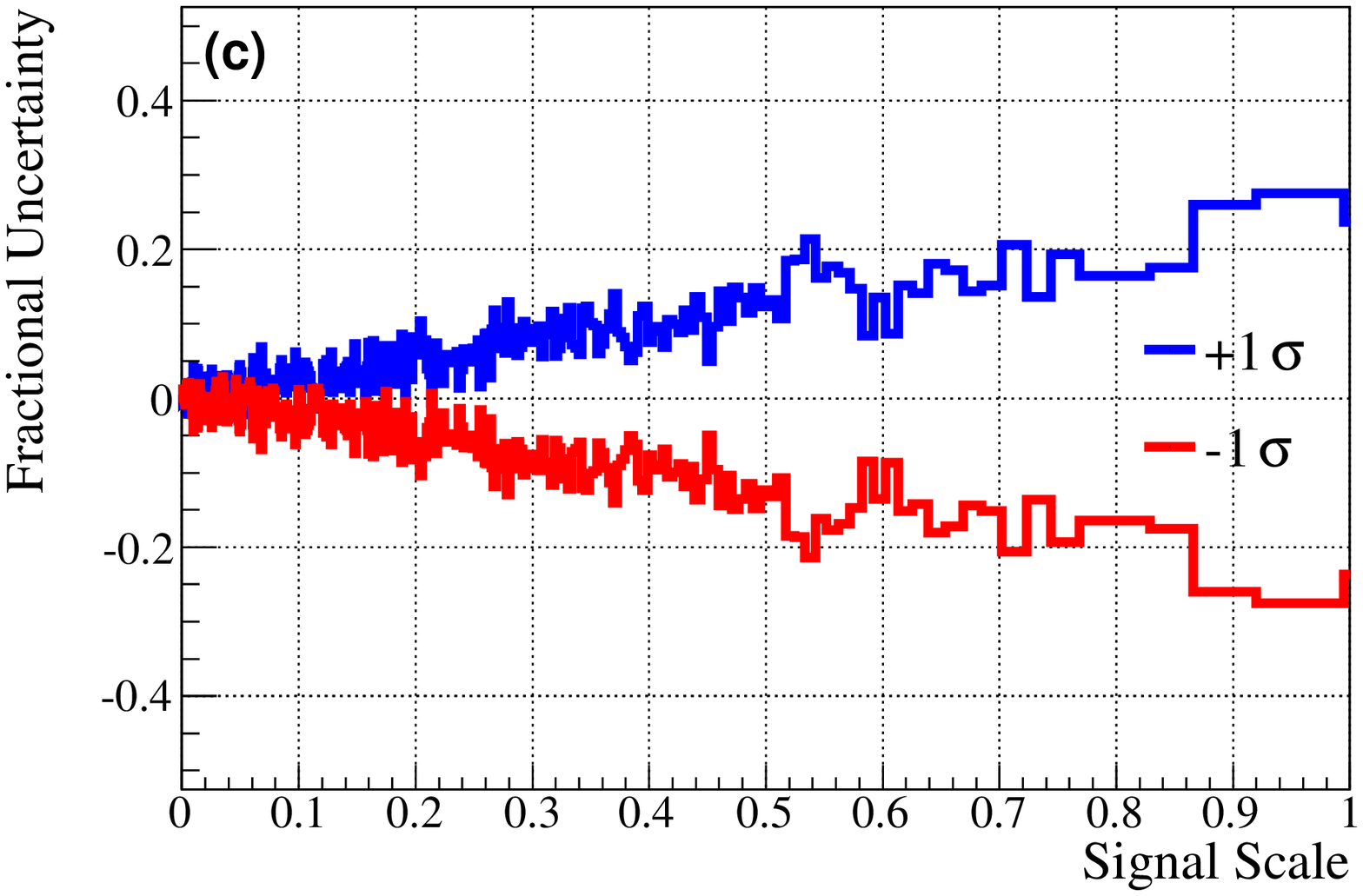}}
\put(8.0,1){\makebox(8,4)[b]{\begin{minipage}[b]{6cm}
\protect\caption{{\footnotesize \small A background-only shape systematic uncertainty with the $\tau$ value is deviated from 0\% to 10\%, for both positive and negative directions. (a) Again 95\% is used for the confidence level requirement and (b) displays the ratio. (c) shows the fractional uncertainty when the $\tau$ value deviates by 5\%.}
\protect\label{fig:TempBaaScSi2}}
\end{minipage}}}
\end{picture}
\label{fig:tau}
\end{figure}
\renewcommand{\baselinestretch}{1}
\small\normalsize

\noindent The AWW approximation stays consistently below the {\tt Collie} value by around 1.5\% and follows the same trend. This test was run with a 5\% rate systematic uncertainty on the background, which holds the ratio maximum at around 0.95. The ratio varies within a percent of 95\%, therefore the approximation is valid.

\subsection{Varying the Number of Histogram Bins}
\label{VaNuBi}

An inherent loss of information occurs when data is binned. Due to this, we want to test the ability of the AWW approximation to reproduce the level of information loss of the full calculation by varying the number of bins. Our results are presented in Fig.~(\ref{fig:histogram}).

\setlength{\unitlength}{1.0 cm}
\renewcommand{\baselinestretch}{0.9}
\begin{figure}[htbp]
\begin{picture}(10.0,5)
\put(0,0)
{\includegraphics{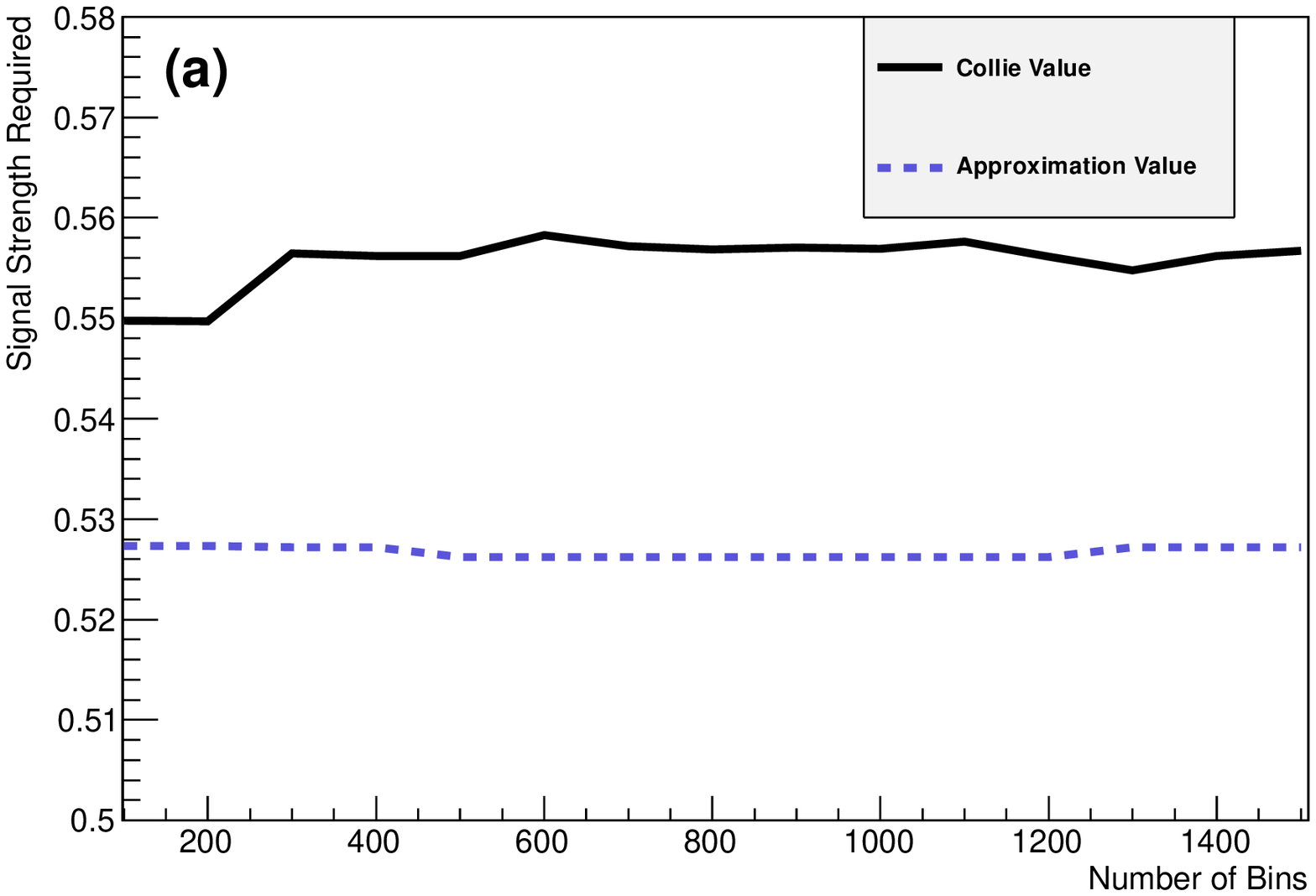}}
\put(8,0)
{\includegraphics{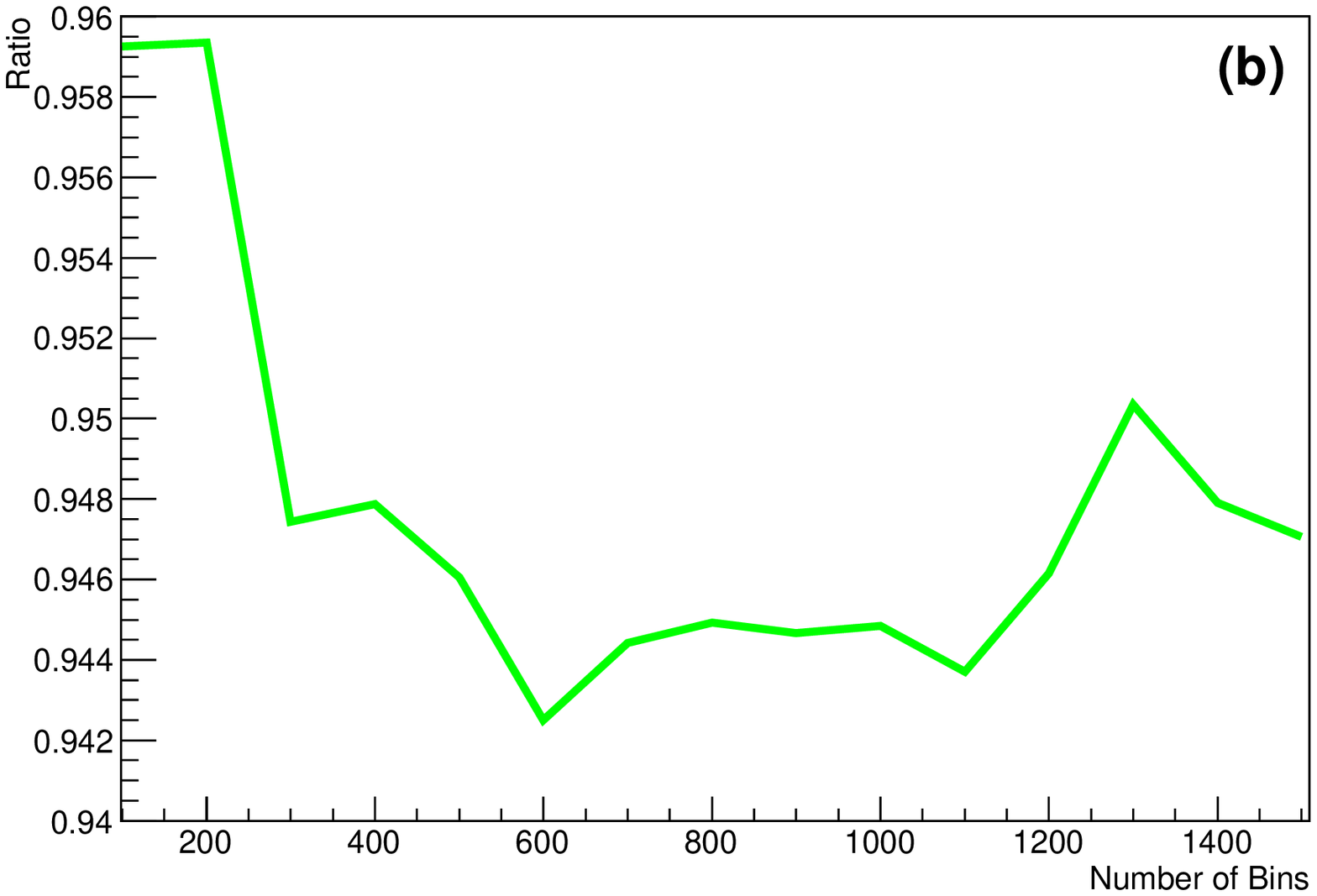}}
\end{picture}
\caption{\small The number of bins was varied from 100 to 1500 in 100 step increments. (a) Displays the signal scale required to achieve 95\% confidence and (b) the ratios for both methods.}
\label{fig:histogram}
\end{figure}
\renewcommand{\baselinestretch}{1}
\small\normalsize

\noindent This test was run with a 5\% background-only rate systematic uncertainty. As is consistent with this additional uncertainty the ratio stays around 95\%; the AWW approximation is valid in reproducing equivalent information loss.

\FloatBarrier

\subsection{Variation in the Number of Events}
\label{sec:Iter}
The last test of the system we built was by varying the number of data used. We wanted to find how many data points were necessary in order to achieve a usable approximation. These results are plotted in Fig.~(\ref{fig:Iter}).

\setlength{\unitlength}{1.0 cm}
\renewcommand{\baselinestretch}{0.9}
\begin{figure}[htbp]
\begin{picture}(10.0,6)
\put(0,0)
{\includegraphics{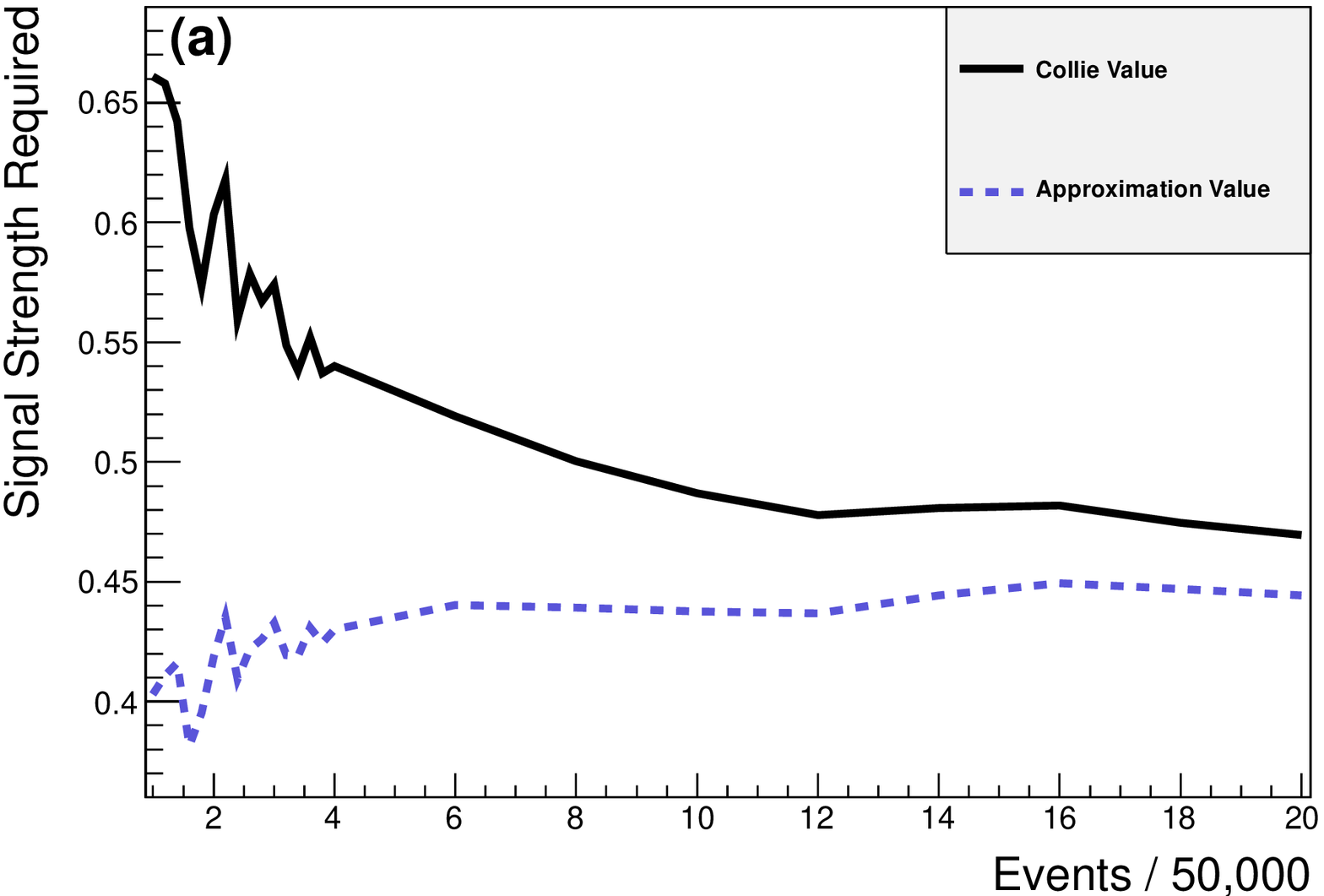}}
\put(8,0)
{\includegraphics{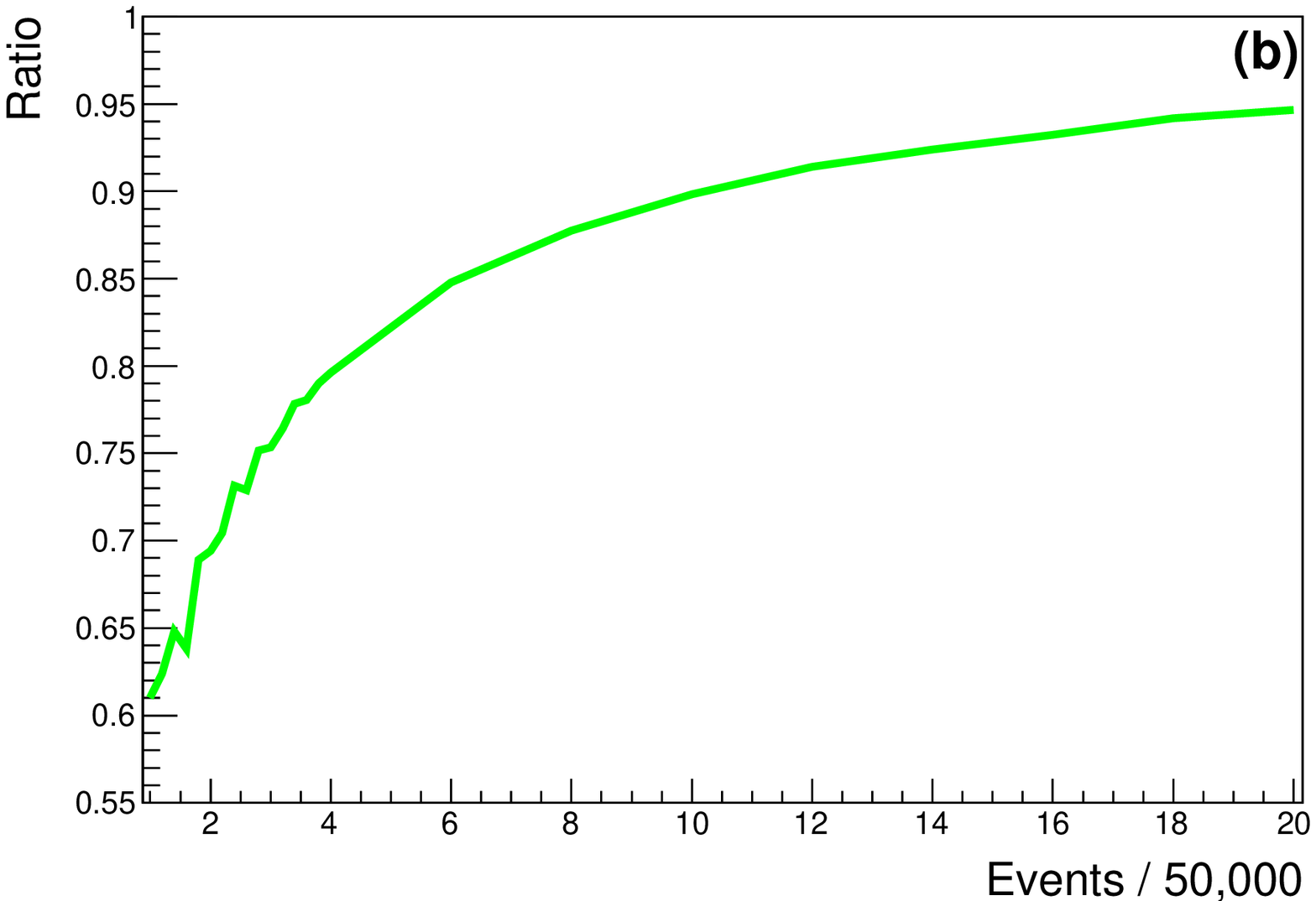}}
\end{picture}
\caption{\small The {\tt Collie} and the approximation values for the signal strength required for 95\% confidence are plotted as a function if the number of data points. The X-axis is labeled as events/50,000. 50,000 is the lowest number of data points that returned usable values for both calculations; the number of iterations is equal to the number of data points. (b) shows the ratio of the two. These plots were generated with a 5\% rate systematic uncertainty on the background.}

\label{fig:Iter}
\end{figure}
\renewcommand{\baselinestretch}{1}
\small\normalsize

\noindent When the number is too small the conditions for Wilks' Theorem are not met, which invalidates the AWW approximation under these conditions. This is evident on the ratio plot, where there is an asymptotic behavior as the number of events increases. This was applied with a 5\% background-only rate systematic uncertainty, so the limit approaches about 0.95.

\section{Conclusion}
\label{sec:Conc}

In summary, we tested the AWW approximation against the full semi-frequentist calculation, with no approximations, as calculated in {\tt Collie}. We ran background-only rate systematic uncertainties, background-only and signal shape systematic uncertainties, asymmetric Gaussian flat systematic uncertainties, varied the background shape itself, varied the number of bins, and varied the number of events. The AWW approximation behaved as expected based on the results from \cite{Asym}.

The tests where the model correctly reproduces the parameter values of the full calculation include the rate systematic uncertainties, the background and signal shape uncertainties, the number of histograms bins, and the uncertainty in the background shape. The shape systematic uncertainties on only background, and the combined shape and background systematic uncertainties run at about 95\% of the true value, i.e. the AWW approximation would exclude with 95\% the signal strength required of the full calculation. When there are no systematic uncertainties the two methods returned nearly equal values. None of the figures for these tests show any absolute trend. 

\noindent The tests where the AWW model breaks down occur where expected. The first of these are the asymmetric Gaussian tests. In the case where the asymmetry is small, roughly at or below 25\% difference (A=2/3), the AWW approximation and {\tt Collie} agree. But when the difference is greater the AWW approximation fails. The second test where the model fails to reproduce the full calculation value is where the number of data points is varied. At low numbers the model fails to reproduce the full calculation, but as the number increases it approches an asymptotic value close to that of the full calculation.

These results are as expected given the two approximations, Wilks and Wald, combined to form the new approximation, the Asimov data set, and is consistent with the report this paper examines. One of the conditions for Wilks' theorem is using a sufficiently large sample and one of the conditions for Wald's theorem is that the data uncertainties follow a Gaussian distribution (There are more conditions necessary to use either theorem, but these are the two that explain the behavior found in tests where the AWW approximation fails). In the case where an asymmetric Gaussian becomes non-Gaussian the model fails, as expected according to Wald's theorem and as the number of data points falls, the mentioned condition for Wilks's theorem fails (as well as increasing the neglected term in the Wald formula).

Therefore, we conclude that when the conditional definitions of Wilks and Wald are met, then the approximation presented in {\it Asymptotic formulae for likelihood-based tests of new physics} does reproduce the full calculation reliably within 5-10\%. Our results suggest that the approximations, published by Cowan, Cranmer, Gross, and Vitells, has the correct asymptotic behavior as designed. Though this approximation has limitations when any of the component approximations are explicitly invalidated, also as expected.

\end{document}